\documentclass[aip, cha, reprint, groupedaddress, showpacs, floatfix]{revtex4-1}
\usepackage{amsmath,graphicx,multirow,natbib,hyperref}
\usepackage{dcolumn}
\usepackage{bm}

\hypersetup{colorlinks,citecolor=black,filecolor=black,linkcolor=black,urlcolor=black}
\usepackage{amsfonts}
\usepackage[utf8]{inputenc}
\usepackage[T1]{fontenc}
\usepackage{amsthm}
\usepackage{amsmath}
\usepackage{amssymb}

\usepackage{epsfig}
\usepackage[percent]{overpic}
\usepackage{pict2e}
\usepackage{psfrag}
\usepackage{verbatim}
\usepackage{color}
\usepackage{placeins}

\usepackage[all]{xy}
\usepackage{enumerate}

\usepackage{tikz}
\usetikzlibrary{matrix}

\usepackage{mathptmx} 

\theoremstyle{plain}

\theoremstyle{definition}

\newtheoremstyle{myremark}
{3pt}
{3pt}
{\small \rmfamily}
{5pt}
{\rmfamily}
{:}
{.5em}
{}

\newcommand{\txtd}{\text{d}}
\newcommand{\R}{\mathbb{R}}
\newcommand{\Z}{\mathbb{Z}}
\newcommand{\AER}{A^\text{ER}}

\newcommand{\xiN}{\xi^{(N)}}

\begin{document}
\title[]{First-order like phase transition induced by quenched coupling disorder}
\author {Hyunsuk \surname{Hong}}
\email{hhong@jbnu.ac.kr}
\affiliation{Department of Physics and Research Institute of Physics and Chemistry, Jeonbuk National University, Jeonju 54896, Korea}
\affiliation{School of Physics, Korea Institute for Advanced Study, Seoul 02455, Korea}
\author{Erik A. Martens}
\email{erik.martens@math.lth.se}
\affiliation{Centre for Mathematical Sciences, S\"{o}lvegatan 18, Lund University, 221 00 Lund, Sweden}
\affiliation{Chair for Network Dynamics, Center for Advancing Electronics Dresden (cfaed) and Institute for Theoretical Physics, TU Dresden, 01062 Dresden, Germany}
\date{\today}

\begin{abstract}
We investigate the collective dynamics of a population of $XY$ model-type oscillators, globally coupled via non-separable interactions that are randomly chosen from a positive or negative value, and subject to thermal noise controlled by temperature $T$. For a finite ratio of positive versus negative coupling, we find that the system at $T=0$ exhibits a discontinuous, first-order like phase transition from the incoherent to the fully coherent state. 
\textcolor{black}{
We determine the critical threshold for this synchronization transition using a linear stability analysis for the fully coherent state and a heuristic stability argument for the incoherent state. 
Our theoretical results are supported by extensive numerical simulations which clearly display a first order like transition. 
Remarkably, the synchronization threshold induced by the type of random coupling considered here is identical to the one found in studies which consider uniform input or output strengths for each oscillator node [Hong and Strogatz, Phys. Rev. E \textbf{84}, 046202 (2011); H. Hong and S. H. Strogatz, Phys. Rev. Lett. \textbf{106}, 054102 (2011)]. 
When thermal noise is present, $T>0$, 
the transition from incoherence to the partial coherence is continuous and the critical threshold is now larger compared to the deterministic case, $T=0$.
We formulate an exact mean-field theory applicable to heterogeneous network structures, based on recent work for the stochastic Kuramoto model using graphon objects representing graphs in the mean-field limit. 
Applying stability results for the incoherent solution, we derive an exact formula for the synchronization threshold for $T > 0$. In the limit $T\rightarrow 0$ we retrieve the result for the deterministic case with $T=0$.
}
\end{abstract}

\pacs{05.45.-a, 89.65.-s}

\keywords{synchronization, coupled oscillators, random coupling, thermal noise, $XY$ model, Kuramoto model, graphon}

\maketitle
\begin{quotation}
The phenomena of magnetization and synchronization occurring in spin models and coupled oscillator models, respectively, are traditionally regarded as completely separate and were therefore mostly been studied independently. However, the two models share collective behaviors induced by the interacting units present in the system. 
Various collective behaviors can be observed depending on the interaction type
such as ``glassy behavior'', where spins become ``frustrated'' when the interaction among spins is chosen randomly from either positive or negative values.
The equations of motion for the spin models is known as the $XY$ model and can be related to a variant of the Kuramoto model of coupled oscillators, an observation which motivates the present study. We considered interactions (coupling strengths) drawn randomly from either a fixed positive or negative value, and studied the resulting collective dynamics of the system using numerical simulations. \textcolor{black}{We find that for the deterministic case, when noise is absent, the system shows features of a discontinuous, first-order like phase transition between incoherent and  perfectly coherent oscillations; for the noisy case, on the other hand, this transition is found to be continuous. We explain and analyze these phase transition using simple energetic arguments, linear stability analysis and a recently developed mean-field theory based on graphons/graphops\cite{ChibaMedvedev2016,Kuehn2020,Gkogkas2022}.}
\end{quotation}

\section{Introduction}
Synchronization is a collective dynamical phenomenon that manifests itself in an impressive range of physical and biological systems~\cite{Strogatz2012,PikovskyRosenblumKurths2001}.
Here we explore the emergent collective behavior exhibited by a population of all-to-all coupled oscillators with random symmetric coupling strengths, either positively or negative valued, and subject to thermal noise.
This system is closely related to the mean-field $XY$ model with random coupling disorder, also known as the spin-glass model~\cite{FischerHertz1993,SherringtonScott1975}; however, interactions in the $XY$ model are asymmetric rather than symmetric, and coupling strengths obey a normal distribution, rather than being randomly drawn from two distinct values.
Furthermore, in the research fields of coupled oscillator networks and synchronization theory much attention has been paid to ``oscillator glasses'', i.e., the possibility of glass-like behavior has been explored by investigating the relaxation dynamics of such systems~\cite{Daido1987,Daido1992,Daido2000,Stiller1998,Stiller2000,Ottino2018,Iatsenko2014}.
In this context we note that for analytical tractability, most studies on oscillator-glasses focused on ``separable'' disorder in the coupling, where the interaction between two oscillators is given by the product of two coupling strengths associated with the adjacent oscillator notes individually, e.g., $K_{ij}=J s_i s_j$ where $s_{i(j)}=\pm 1$ and $J$ = const.

In this study, we were instead interested in the behavior induced by non-separable symmetric coupling interactions between oscillator nodes whose interaction strengths were randomly chosen from either of two values, one negative ($K_n<0$) and positive ($K_p>0$). In exploring the collective behavior of the system we focused on determining the critical threshold for the transition from incoherent to coherent oscillatory behavior; and especially in the possibility of a {\it universal character} regarding  models with varying types of coupling interactions. Specifically, we compared the synchronization threshold of the present model (random non-separable coupling interactions) with the one found for models studied previously by one of the authors, where non-separable interactions are characterized by uniform input or output strengths between oscillator nodes~\cite{HongPRL2011,HongPRE2011,Hong2012}. In doing so, we considered the effects of quenched random coupling disorder and of thermal noise whose strength is controlled by a temperature $T$.

The paper is structured as follows.  In Sec.~II we introduce the model with its governing equations. Sec.~III discusses the emergent collective behavior of the system for zero noise ($T=0$) and explains the phase transition from the incoherent to coherent oscillator state.  We further investigate this transition via linear stability analysis.  In Sec.~IV, we explore the effect of non-zero thermal noise ($T>0$) on the collective behavior of the system, based on numerical simulations and an exact mean-field theory. The article is concluded with a brief summary and discussion in Sec.~V.

\section{Model}
We consider $N$ coupled oscillators 
$i=1,\ldots,N=:[N]$ whose phases $\theta_i(t) \in \R / 2\pi \Z$ obey the dynamics given by
\begin{equation}
    \frac{\txtd \theta_i}{\txtd t} = \frac{1}{N}\sum_{j =1}^N K_{ij} \sin(\theta_j-\theta_i) + \eta_i(t),
\label{eq:model} 
\end{equation}
where $K_{ij}$ is the coupling strength between oscillator nodes $i$ and $j$, drawn from the bimodal distribution function    
\begin{equation}
    h(K)=p\,\delta(K-K_p)+(1-p)\,\delta\,(K-K_n),
    \label{eq:hK}
\end{equation}
where positive ($K_{p} > 0$) and negative ($K_{n} < 0$) coupling occur with probability $p$ and $1-p$, respectively. 
We require that the interactions be symmetric between oscillator nodes, i.e., $K_{ij}=K_{ji}$ for all $i$ and $j$; the model definition implies vanishing self-interaction which is equivalent to letting $K_{ii} = 0$, for all oscillators $i\in[N]$. 
Rescaling time, $t\mapsto K_p \,t$, we may reduce the number of effective system parameters by introducing the positive valued \emph{coupling ratio} $Q:= -K_{n}/K_{p} >0$. For now, we shall refer to the unscaled coupling, $K_{ij}$, to complete a physical heuristic discussion until Sec.~\ref{lab:phyicalargument}; from  Sec.~\ref{lab:stability_coherence} onwards, we shall make use only of the rescaled version of the coupling to perform various analysis (i.e. let $K_p =1 $ and $K_n=-Q$).

Note that it is possible to write out the coupling matrix explicitly, in terms of  a linear combination of a constant and the Erdösz-Rényi (ER) network, which will prove useful in the subsequent analysis:
\begin{align}\label{eq:Kmatrix}
    \begin{split}
        K_{ij} &= K_n + (K_p - K_n) \, \AER_{ij}\\
        &= K_p(-Q + (1 + Q) \, \AER_{ij}\
    \end{split}
\end{align}
where the ER network is defined via the symmetric adjacency matrix $\{\AER_{ij}\}_{i,j\in[N]}$ with $\AER_{ii}=0, i\in[N]$, taking on the value 1 with probability $p$  and 0 otherwise; i.e., let $\AER_{ij}=\AER_{ji}$ be independent Bernoulli variables, $\mathbb{P}(\AER_{ij})=p$ with $0<p<1$ for $1\leq i<j\leq N$, and $\AER_{ij} = 0$ otherwise.

The presence of attractive ($K_{ij}>0$) versus repulsive coupling ($K_{ij}<0$) side-by-side may seem unusual for physical systems, but they are common in the context of excitatory/inhibitory interactions occurring (neuro-) biological systems~\cite{VreeswijkSompolinsky1996,BorgersKopell2003} and of 
social systems~\cite{Galam2004} consisting of agents with conformist and contrarian behavior~\cite{HongPRL2011,HongPRE2011,Hong2012}; however, while those studies considered coupling with \emph{uniform input strength} ($K_{ij}=K_i$ for all $i$) or \emph{uniform output strength} ($K_{ij}=K_j$  for all $j$) from each oscillator node, here we are concerned with non-uniform (but symmetric) interactions, similar to spin glass models~\cite{FischerHertz1993,SherringtonScott1975}.
The last term $\eta_i(t)$ in Eq.~\eqref{eq:model} models thermal noise in the system, i.e., mean-centered $(\langle \xi_i(t) \rangle=0)$  Gaussian noise with variance 
\begin{equation}
    \langle \eta_i(t) \eta_j(t^{\prime}) \rangle = 2T\delta_{ij}\delta(t-t^{\prime}),
    \label{eq:thermalnoise}
\end{equation}
where $T$ defines the temperature via the Einstein-Smoluchowski relation
$D=k_B T/\gamma$ where $k_B$ is the Boltzmann constant $k_B$ and we set the fraction coefficient to $\gamma = 1$. 

The presence of both attractive and repulsive interaction  induces  {\it{frustration}} in the oscillator system. In this context, we note that Eq.~\eqref{eq:model} is related to the mean-field $XY$ spin-glass model~\cite{SherringtonScott1975}
with {\it{random}} coupling disorder. Introducing the Hamiltonian 
\begin{equation} 
{\cal{H}}=-\frac{1}{2N}\;\sum_{i=1}^N\; \sum_{\substack{j=1,\\(j\neq i)\;}}^NK_{ij}\cos(\theta_j - \theta_i),
\label{eq:Hamiltonian}
\end{equation}
the Langevin dynamics in Eq.~\eqref{eq:model} can expressed as
$\dot\theta_i = -\frac{\partial {\cal{H}}}{\partial \theta_i}+\eta_i(t)$.
This formulation allows the interpretation of the model as the ``overdamped" version of the Hamiltonian dynamics at finite temperature $T$, i.e., the {\it{inertia}} term proportional to ${\ddot{\theta}}_i$ is negligible compared to the damping term proportional to $\dot{\theta}_i$.

\section{Phase transition at $T=0$}
To examine how random coupling affects the collective behavior,
we first explore the case of zero temperature ($T=0$) to understand which collective states are possible in the system \eqref{eq:model}.
To gain such insight, consider two extreme cases in $K_{ij}$ in Eq.~\eqref{eq:hK}: 
\begin{itemize}
 \item [(i)] For $p=1$ we have uniform coupling with $K_{ij} = K_p (>0)$ for all $i$ and $j$, 
which can be interpreted as ferromagnetic coupling. The only energetically favorable state of the system is then fully coherent (ordered), since it possesses the lowest energy with identical phase angles ($\theta_i = \theta_j$ for all $i$ and $j$). 
\item [(ii)] For $p=0$, the coupling is uniform with antiferromagnetic character, $K_{ij}=K_n (<0)$ 
for all $i$ and $j$. The system then exhibits the incoherent state with random phases $\theta_i \in [0, 2\pi)$, which corresponds to a zero energy in the average sense as $N\rightarrow \infty$.
\end{itemize}
In between these two cases, we expect a phase transition from 
the incoherent state to the fully coherent state at a certain critical value $p=p_c$. This hypothesis is corroborated by the following two theoretical arguments which predict the critical threshold, $p_c$, for the transition from incoherence to coherence (disorder to order).

\subsection{Physical argument}\label{lab:phyicalargument}
The transition point $p_c$ can be directly deduced from the properties of the 
distribution  $h$. Given that the random coupling $K_{ij}$ is the only source of disorder in the system, we expect that the critical $p_c$ is controlled by its mean value, $\langle K_{ij} \rangle = p K_{p} + (1-p) K_{n}$ for the distribution given by Eq.~\eqref{eq:hK}. Specifically, we expect the transition to occur when $\langle K_{ij} \rangle $ changes sign, i.e., when the coupling switches from being predominantly negative to being predominantly positive. This is the case when $\langle K_{ij} \rangle=0$, which implies that 
\begin{align}\label{eq:pc_zeroT}
    p_c&=\frac{Q}{1+Q} .\
\end{align}
The plausibility of this result becomes evident when considering the special case $Q=1$, where negative and positive coupling is equally distributed: attractive coupling dominates over repulsive coupling exactly when $p_c=1/2$ and thus  the onset of coherence is induced. 

Alternatively, the threshold $p_c$ can be obtained by an energetic argument. For the fully coherent state, the $XY$ oscillators possess the same phases, 
$\theta_i(t) = \theta_j(t)$ for all $i$ and $j$, which implies that $R=1$ in Eq.~\eqref{eq:R}. 
Accordingly, the energy density $\epsilon:=\mathcal{H}/N_\text{tot}$ (energy per spin/oscillator) for the fully 
coherent state in an average sense can be stated as
\begin{eqnarray}
{\cal{E}}_\text{coh} &=& -[K_p (N_p/N_{\rm{tot}}) + K_{n}(N_{n}/N_{\rm{tot}})], 
\end{eqnarray}
where $N_{p}$ (or $N_n$) is the number of positive (or negative) couplings 
among the total number of couplings $N_{\rm{tot}}=N(N-1)/2$; in the thermodynamic limit ($N\rightarrow \infty$) the energy density becomes
\begin{align}
    {\cal{E}}_\text{coh}&= -[K_{p} \; p + K_{n} (1-p)]. 
\end{align}
It is clear that by definition, $\mathcal{E}_\text{coh}\leq 0$ for all values of $0 \leq p\leq 1$ and $Q>0$.
Note that for this special case of the fully coherent state, the energy relates directly to the average coupling strength, $\mathcal{E}_\text{coh}=-\langle K_{ij}\rangle$.

Similarly, the energy density for the incoherent state in the thermodynamic limit can be expressed as 
\begin{equation}
    {\cal{E}}_\text{inc} = -[K_{p}\, p\, \xi + K_{n}\,(1-p) \,\xi^{\prime}], 
\end{equation}
where the (ensemble) averages of phase differences are defined as $\xi := \langle \cos(\theta_i-\theta_j) \rangle$ and $\xi':= \langle \cos(\theta_i'-\theta_j') \rangle$. Since all oscillators in the incoherent state assume arbitrary phase values over time with equal probability, we have that $\xi=\xi'=0$, and therefore ${\cal{E}}_\text{inc}=0$ for $N\rightarrow\infty$.
At the transition ($p=p_c$), continuity of the energy density demands that the two energy densities match, $\mathcal{E}_\text{coh}=\mathcal{E}_\text{inc}$, which implies $(K_{p} - K_{n})p_c+K_{n} = 0$. As a consequence, we obtain the same result as in~\eqref{eq:pc_zeroT}.

\begin{figure}
\includegraphics[width=\columnwidth]{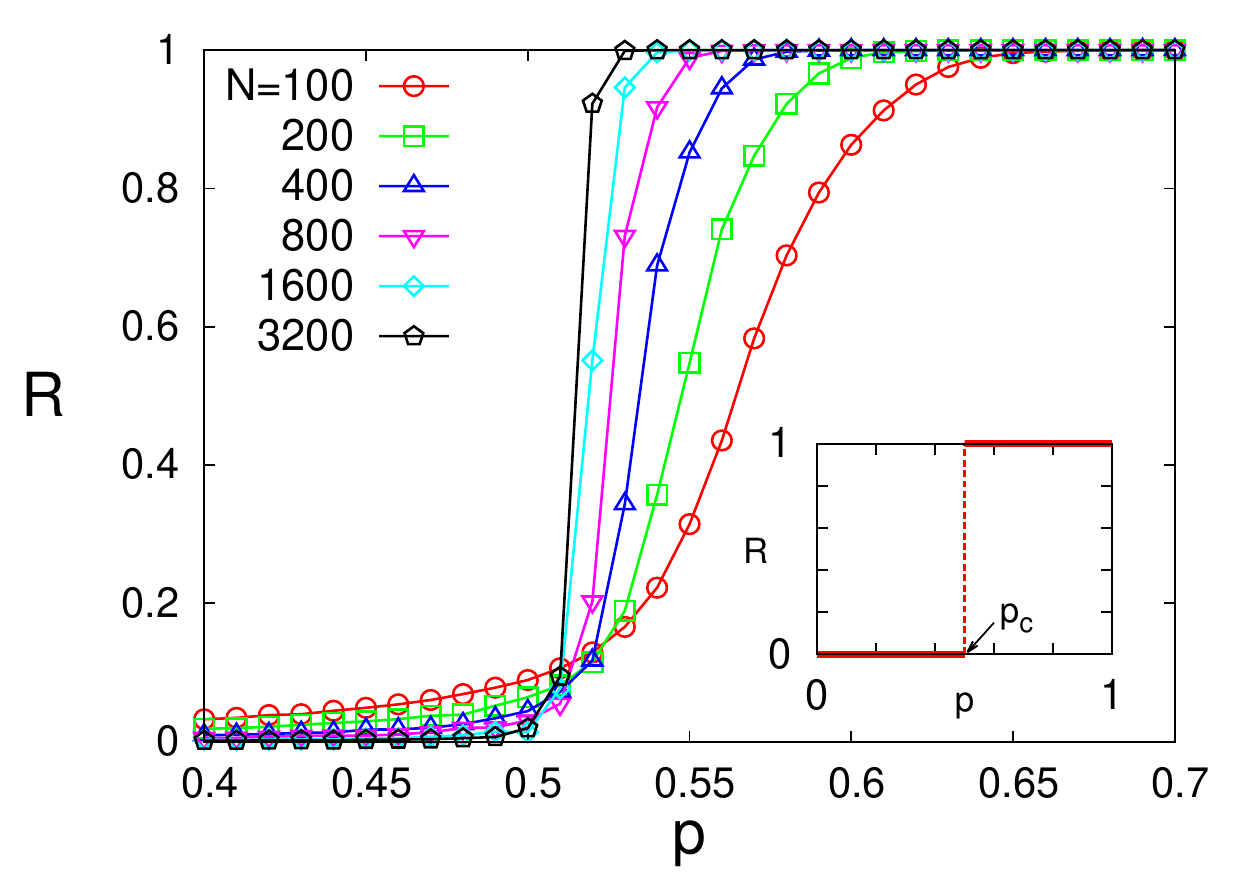}
\caption{(Color Online) The behavior of $R$ is shown in dependence of $p$ for $Q=1$ and $T=0$ for varying the system sizes $N$.  
The slope of $R$, i.e., $\txtd R/\txtd p$ increases as the size $N$ 
is augmented, suggesting a first-order like transition 
near the critical threshold  $p_c=1/2$  predicted for $Q=1$ using ~\eqref{eq:pc_zeroT}.
The inset shows a behavior of $R$ as expected for the thermodynamic 
limit, $N \rightarrow \infty$ (the dashed line is a guide to the eye). 
}
\label{fig:R}
\end{figure}
To see that such phase transition is indeed possible, and to confirm our theoretical prediction for the critical threshold \eqref{eq:pc_zeroT}, we numerically integrated Eqs.~\eqref{eq:model} 
for the time $M_t=2\times 10^5$ using the Heun method~\cite{Heunmethod} with discrete time increments of $\Delta t=0.01$. Initial phases  were drawn from the i.i.d. uniform distribution on $\{\theta_i(0)\}_{i\in[N]} \in [-\pi,\pi)$.  Transient behavior possibly occurring during the initial time $t<3/4 M_t =1.5\times 10^5$ was discarded until an equilibrium state was reached, and observed quantities were averaged over the remaining simulation time.  
We measured the collective behavior of the system using the complex order parameter
\begin{equation}
R = \Bigg\langle \Bigg| \frac{1}{N} \sum_{j=1}^N e^{i\theta_j} \Bigg|\Bigg\rangle_t, 
\label{eq:R}
\end{equation}
where $|\cdot|$ denotes the absolute value and $\langle \cdot \rangle_t $ the average over the time interval $3/4 M_t \leq  t \leq M_t$. 
In the  framework of the $XY$ spin model~\eqref{eq:Hamiltonian}, the quantity $R$ corresponds to the {\it{magnetization}} of the material for any given spin configuration.

Fig.~\ref{fig:R} displays the behavior of $R$ as a function of 
$p$ with $Q=1$ fixed while varying the system size $N$.
We observe a phase transition from the incoherent state near $R=0$ to the fully coherent state near $R=1$. Our simulations indicate a critical transition  at $p=p_c=1/2$, which agrees with our theoretical arguments further above~\eqref{eq:pc_zeroT}.
The slope of the curve $R=R(p)$ increases for larger system sizes near the transition.  
This suggests the presence of a first-order like transition at the threshold $p_c$, as one would expect for the thermodynamic limit $N\rightarrow \infty$ based on the argument made further above (see Eq.~\eqref{eq:pc_zeroT} and inset in Fig.~\ref{fig:R}). The absence of a perfectly discontinuous transition, characteristic of a first order transition, is explained by finite size effects and transient dynamics near the transition that smooth the transition.
 
\begin{figure}
\includegraphics[width=\columnwidth]{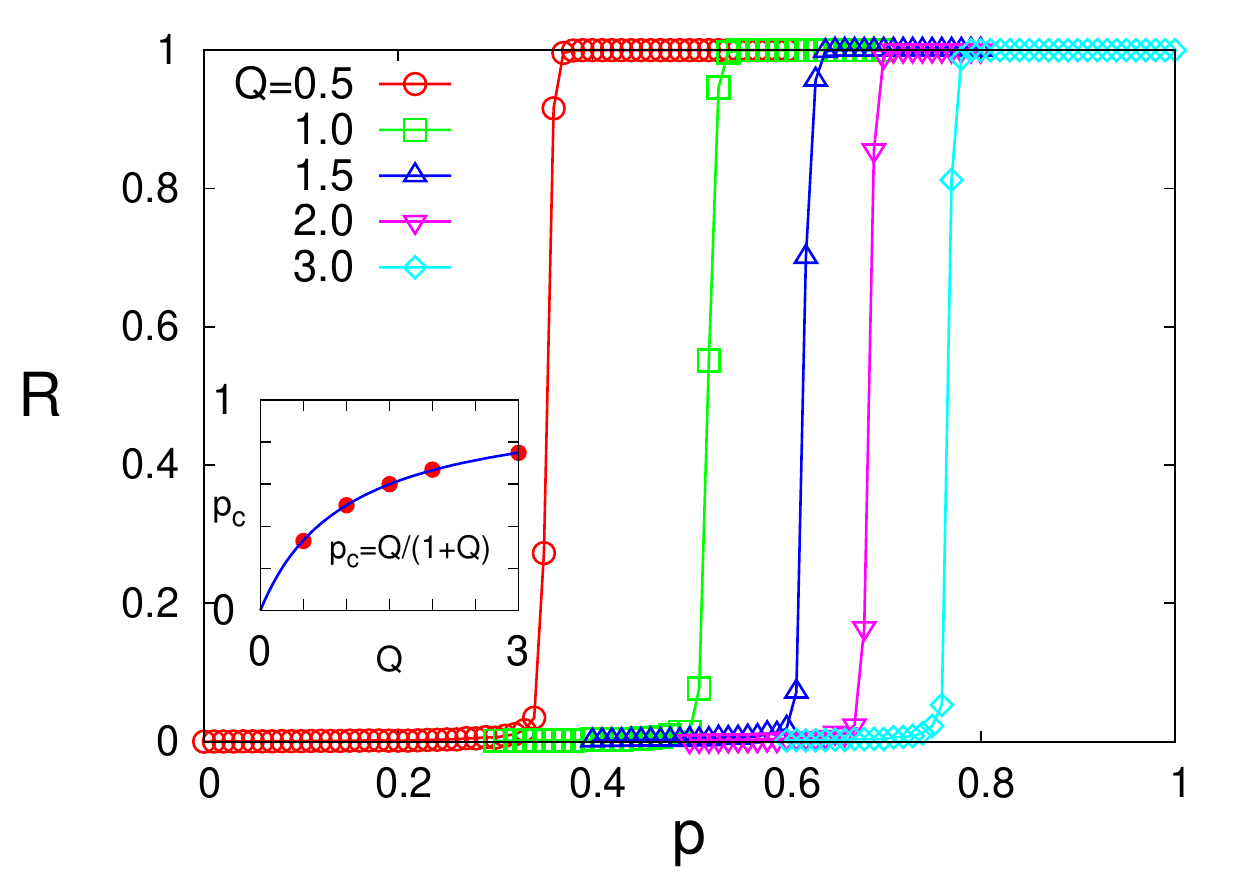}
\caption{(Color Online) Steady state behavior of $R$  in dependence of $p$ for varying coupling ratio $Q$ and $T=0$.  The critical threshold $p_c$ for the phase transition from $R=0$ to $R=1$ increases with larger $Q$ (see inset). 
The system size is set to $N=1600$. Results represent ensemble averages over 10 realizations using different sets of random initial conditions $\{\phi_i(0)\}_{i\in [N]}\in[-\pi,\pi)$ and random coupling strengths $\{K_{ij}\}_{i,j\in[N]}\in\{1,Q\}$ with $Q=-K_n/K_p>0$ (see text). 
The open symbols represent numerical data and lines are guides to the eyes.
Inset: Critical threshold $p_c$ as a function of $Q$. The solid line is given by theoretical prediction in Eq.\eqref{eq:pc_zeroT}. 
}
\label{fig:R_p_variousQ}
\end{figure}

The behavior of $R$ as a function of $Q$ is studied in Fig.~\ref{fig:R_p_variousQ}. We observed that the critical value $p_c$ grows with increasing coupling ratio $Q$.
Augmenting the value of $Q$ increases the ratio of the negative coupling strength $K_n$ versus positive coupling strength $K_p$. It is clear that, as a result, a larger value of $p$ is required to reach the fully coherent state.
We numerically estimated the critical value for $p_c$ based on the simulations using two methods: 
(i) We measured the value $p=p_c$ for which $R(t) > \epsilon$, where we chose $\epsilon=0.02$. 
(Since the value of $R$ for the incoherent state goes to zero according to ${\cal{O}}(N^{-1/2})$, and we measured the $R$ for the system with size $N=1600$ for which $1/\sqrt{1600}=0.025={\cal{O}}(10^{-2})$, the chosen threshold value of $\epsilon=0.02$ is appropriate.). 
(ii) We estimated the critical value $p_c=\text{argmax}_p{(dR/dp)}$ for which the slope of $R$ is maximized.
We found that the values of $p_c$ measured using both methods yield the same result and very well match our theoretical prediction Eq.~\eqref{eq:pc_zeroT} that we derived above, $p_c=Q/(1+Q)$,  as is seen in the inset of Fig.~\ref{fig:R_p_variousQ}.

\subsection{Linear stability analysis of the fully coherent state ($T=0$)}\label{lab:stability_coherence}
The fully coherent state ($R=1$) is characterized by solutions on the coherence (or synchronization) manifold defined by $\theta_i(t)=\theta_j(t)=\theta_s(t)$ for all oscillators $i$ and $j$. Due to the phase-shift invariance, $\theta_i(t)\mapsto \theta_i(t)+\theta_0$, inherent to the model~\eqref{eq:model}, there exists a reference frame co-rotating with the coherent solution in which $\dot{\theta_i}=0$ is satisfied for all oscillators $i$ and $j$. Applying perturbations $\epsilon_i(t)=\theta_i(t)-\theta_s(t)$ to the coherence manifold and linearizing Eq.~\eqref{eq:model}, we obtain the variational equation 
\begin{align}
    \frac{\txtd \epsilon_i}{\txtd t} &= \frac{1}{N}\sum_{j =1}^N K_{ij} (\epsilon_j-\epsilon_i)
    =\frac{1}{N}\sum_{j =1}^N L_{ij} \epsilon_j\
    \label{eq:epsilon}
\end{align}
which is rewritten using the graph Laplacian $L_{ij} = K_{ij}-\delta_{ij}\alpha_i$ with 
Kronecker symbol $\delta_{ij}$ and row sum $ \alpha_i:=\sum_{j=1,j\neq i}^N K_{ij}$.
\begin{figure}
    \includegraphics[width=\columnwidth]{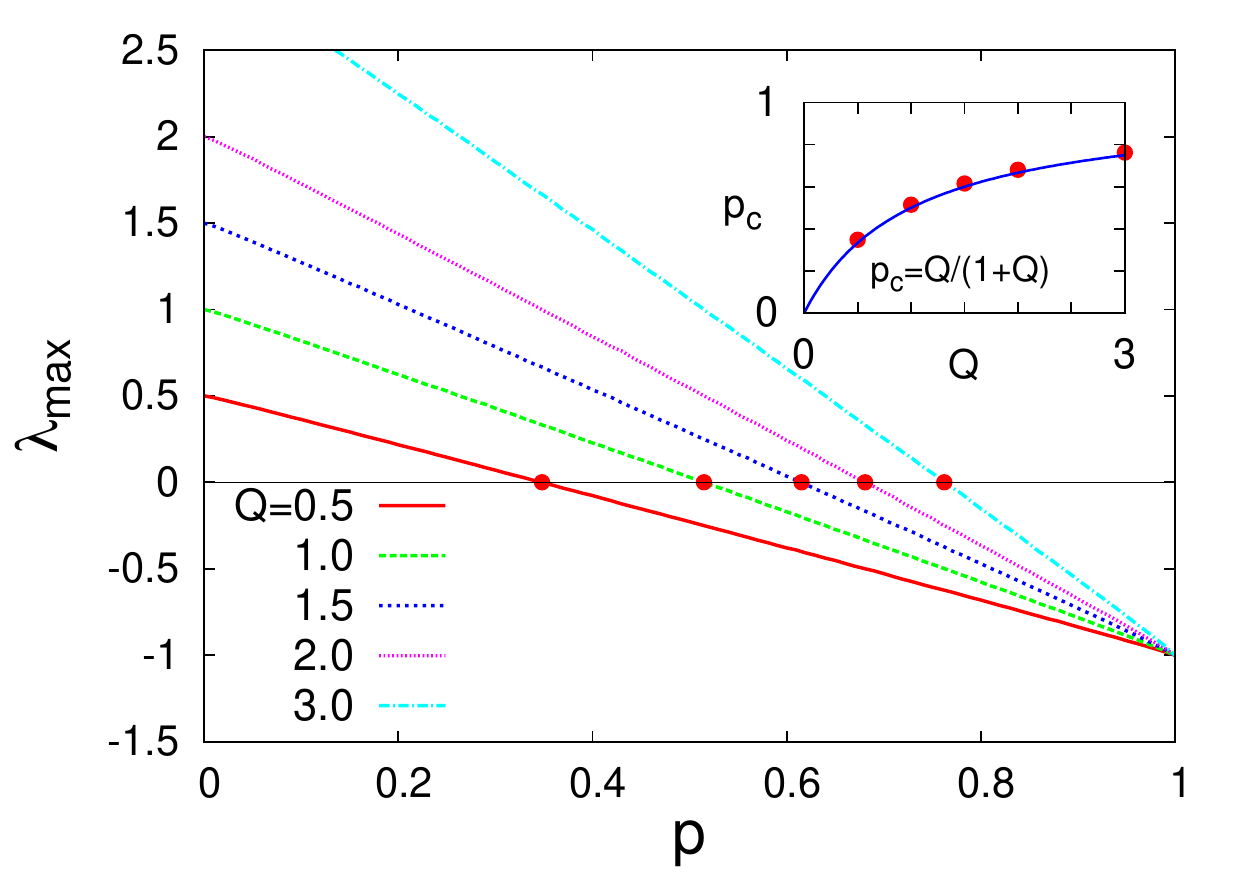}
    \caption{(Color Online) Maximum eigenvalues $\lambda_{\rm{max}}$ shown as a function of $p$ for varying  coupling ratio $Q$ and $T=0$.  Eigenvalues were computed for $N=20'000$ and  averaged over an ensemble of 10 realizations.
    Inset: $p_c$ vs. $Q$ is shown, where the blue solid line is the theoretical prediction $p_c=Q/(1+Q)$.
    }
    \label{fig:Lambdamax_pc}
\end{figure}
The row sum of the Laplacian matrix is $\sum_{j=1}^N L_{ij}=0$, which implies the trivial eigenvector $(1,1,\ldots,1)$ with eigenvalue $\lambda_1=0$; this is a consequence of the phase-shift invariance inherent to the system \eqref{eq:model}.  
Since the Laplacian matrix $L$ is real and symmetric, linear stability for the fully coherent (synchronized) state is guaranteed if all remaining 
$N-1$ (real) eigenvalues are less than zero ($\lambda_i\leq 0$ for $ i=2,3,\ldots,N$).
 
With this idea in mind, we computed the maximum eigenvalue $\lambda_\text{max}<0$ of 
${L}$ for a given value of $Q$, in order to determine the critical $p_c$ for which linear stability of the coherent state breaks down, i.e., $\lambda_\text{max}(p_c)=0$.  Note that for the same value of $Q$, many realizations for $\{K_{ij}\}$ and 
thus also for $\{L_{ij}\}$ can be chosen. 
We therefore computed an average value for $\lambda_{\rm{max}}$ obtained by 10 distinct realizations of $\{L_{ij}\}$. 
Results for $\lambda_{\rm{max}}$ as a function of $p$ and varying values of $Q$ are shown in Figure~\ref{fig:Lambdamax_pc}.
We found that the agreement of these estimates for $p_c$ with the theoretical prediction of $p_c=Q/(1+Q)$ improves as the system size $N$ is increased.

\subsection{Stability argument for the incoherent state ($T=0$)}

In the following, we discuss a heuristic argument to determine the stability of the incoherent state along the line of Winfree's and Kuramoto's original work for all-to-all coupling with uniform strength (see Ref.~\onlinecite{Winfree1967} or \onlinecite{Strogatz2000}, Sections 3 and 4). 
Introducing the weighted order parameter (i.e., a mean-field),
\begin{align}\label{eq:weightedOP}
 r_i e^{i\Phi_i} &:= \frac{1}{N}\sum_{j=1}^N K_{ij} e^{i\theta_j},
\end{align}
where $K_{ij}$ is given by \eqref{eq:Kmatrix},
we can reformulate this field as follows: substituting \eqref{eq:Kmatrix} into \eqref{eq:weightedOP}, and using the definition for the Kuramoto order parameter~\eqref{eq:R} and $Q=-K_n/K_p$ the field is now rewritten as 
\begin{align}\label{eq:weightedOP2}
 r_i e^{i\Phi_i} &= 
-  Q R e^{i\Theta} +  (1+Q)\frac{1}{N}\sum_{j=1}^N \AER_{ij} e^{i\theta_j}
\end{align}
Note that we set $K_p=1$ and $K_n=-Q$ henceforth as mentioned further above.
According to the mean-field theory on annealed networks~\cite{Um2014}, we approximate the ER network as follows, 
\begin{align}
 \AER_{ij} \approx \frac{k_i k_j}{N \langle k \rangle},
\end{align}
where $k_i:=\sum_{j=1}^N \AER_{ij}$ is the degree of the $i$th node (or $j$th node, respectively); and $\langle k \rangle:= \sum_{k=1}^N P_k\, k$ is the mean degree with distribution $P_k$. 
We may approximate the expectation value for the degree, $\langle k \rangle \approx N^{-1}\sum_{j=1}^N k_j = N^{-1} N_e$  where $N_e=N(N-1)p$ is the expected number of edges in $\AER=\AER(p)$.
For $N$ large, we can further approximate $k_i \approx k_j \approx \langle k \rangle \approx N p$, so that 
\begin{align}\label{eq:ER_mf}
 \AER_{ij} &\approx p.\
\end{align}
(Note that this result coincides with the graphon for the ER network, as we discuss Sec.~IV.A. further below)
Using this approximation in \eqref{eq:weightedOP2}, we have
\begin{align}
 r_ie^{i\Phi_i} &= -  Q Re^{i\Theta}+ (1+Q)p R e^{i\Theta}.\
\end{align}
Finally, we formulate the governing equations \eqref{eq:model} in terms of the mean weighted field $r_ie^{\Phi_i}$, and we have $\dot{\theta}_i = r_i \sin{(\Phi_i-\theta_i)}$, i.e.,
\begin{align}\label{eq:KM_weightedOP}
\begin{split}
\dot{\theta}_i &=  (p(1+Q)-Q) R\sin{(\Theta-\theta_i)}.
\end{split}
\end{align}
Expressing the phase dynamics in terms of the order parameter makes the mean-field character of the model obvious. We may now invoke a heuristic argument for the stability of incoherent oscillations~\cite{Strogatz2000}. 

Every oscillator appears to be decoupled from any other, although they all are interacting via the mean-field quantities $R$ and $\Theta$. 
Suppose there exists a quasi-stationary solution (incoherent or coherent) such that the order parameter is constant, $R\exp{(i\Theta)}=\text{const.}$ (this may be checked numerically, or can be proven for the Kuramoto model with uniform coupling strengths). 
Then an oscillator $i$ may be attracted to the angle of the mean-field, rather than to the phase of any other oscillator, i.e., $|\Theta(t)-\theta_i(t)|\rightarrow 0$ as $t\rightarrow \infty$, if 
$(p(1+Q)-Q)>0$
; but not otherwise\footnote{ 
Note also that the effective strength of the coupling is proportional to the coherence $R$. Thus, there is a positive feedback loop between coupling and coherence: as the population becomes more coherent, $R$ grows and so the effective coupling 
$(p(1+Q)-Q)\,R$ 
increases, which in turn may recruit even more oscillators into the synchronized bunch. This process will continue if the coherence is further increased by more oscillators being recruited to the bunch; otherwise, the process becomes self-limiting. }.
Therefore, the critical value where the incoherent state loses stability is expected to occur at
\begin{align}\label{eq:pc_incoherent_T0}
 p_c &= 
 \frac{Q}{1+Q}.\
\end{align}
Note that this value is identical to the critical value where the coherent state loses stability, Eq.~\eqref{eq:pc_zeroT}. This suggests that hysteretic effects cannot be present, which is also confirmed by numerical computation.

\section{Non-zero temperature $(T > 0)$}

Apart from disorder in the coupling, we also investigated how thermal noise in Eqs.~(\ref{eq:model}) affects the collective behavior of the system. Note that the two types of disorder influence the dynamics in different ways: while the  disorder in the coupling, $K_{ij}$, remains constant in time (``quenched'' disorder), the thermal noise ($\eta_i(t)$) models temporal fluctuations. For $T=0$, we found that the system displays a discontinuous phase transition at a critical threshold $p_c=Q/(1+Q)$. How does thermal noise alter the characteristic of the phase 
transition, and what is the critical threshold for $T>0$?

\begin{figure}
\begin{overpic}[width=\columnwidth,percent]{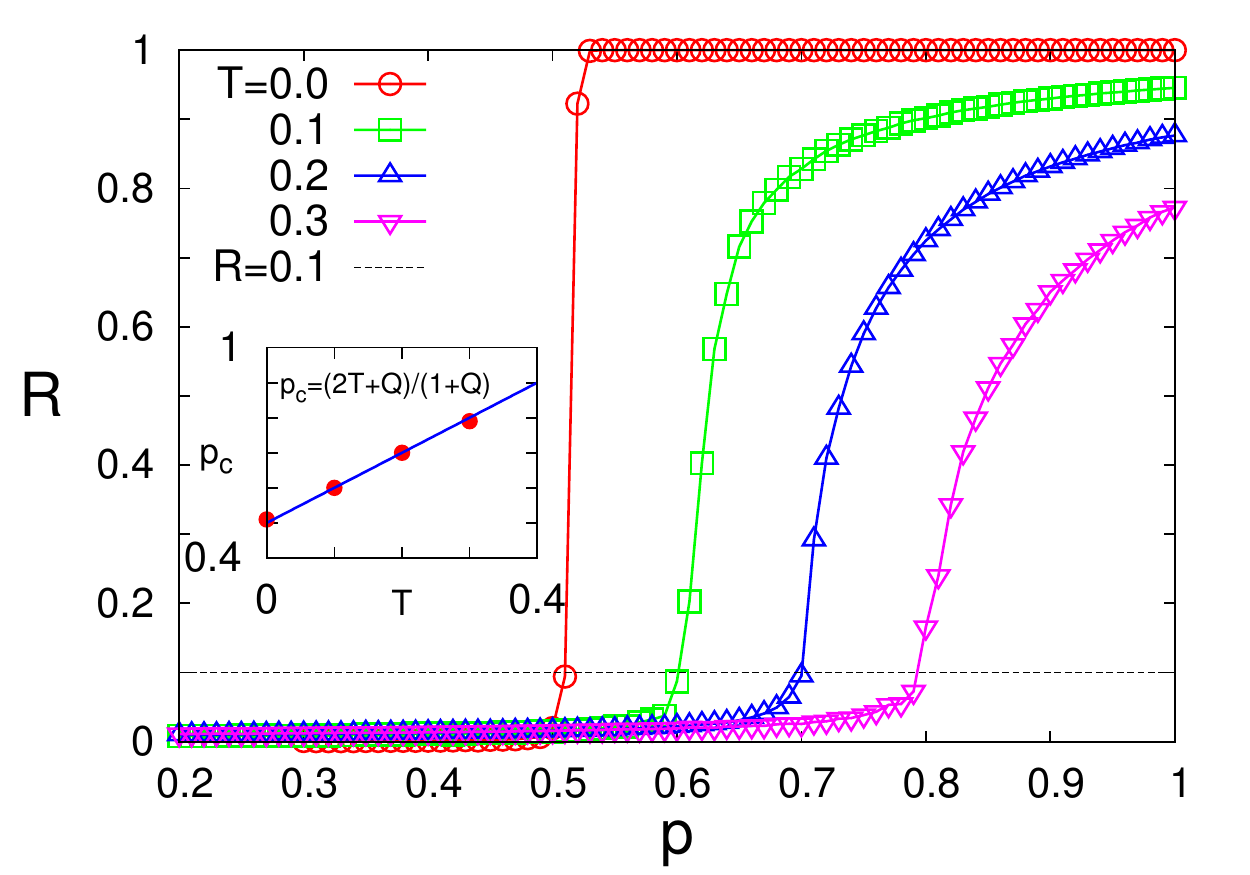}
        \put(0,65){\color{black}(a)}
\end{overpic}
\begin{overpic}[width=\columnwidth,percent]{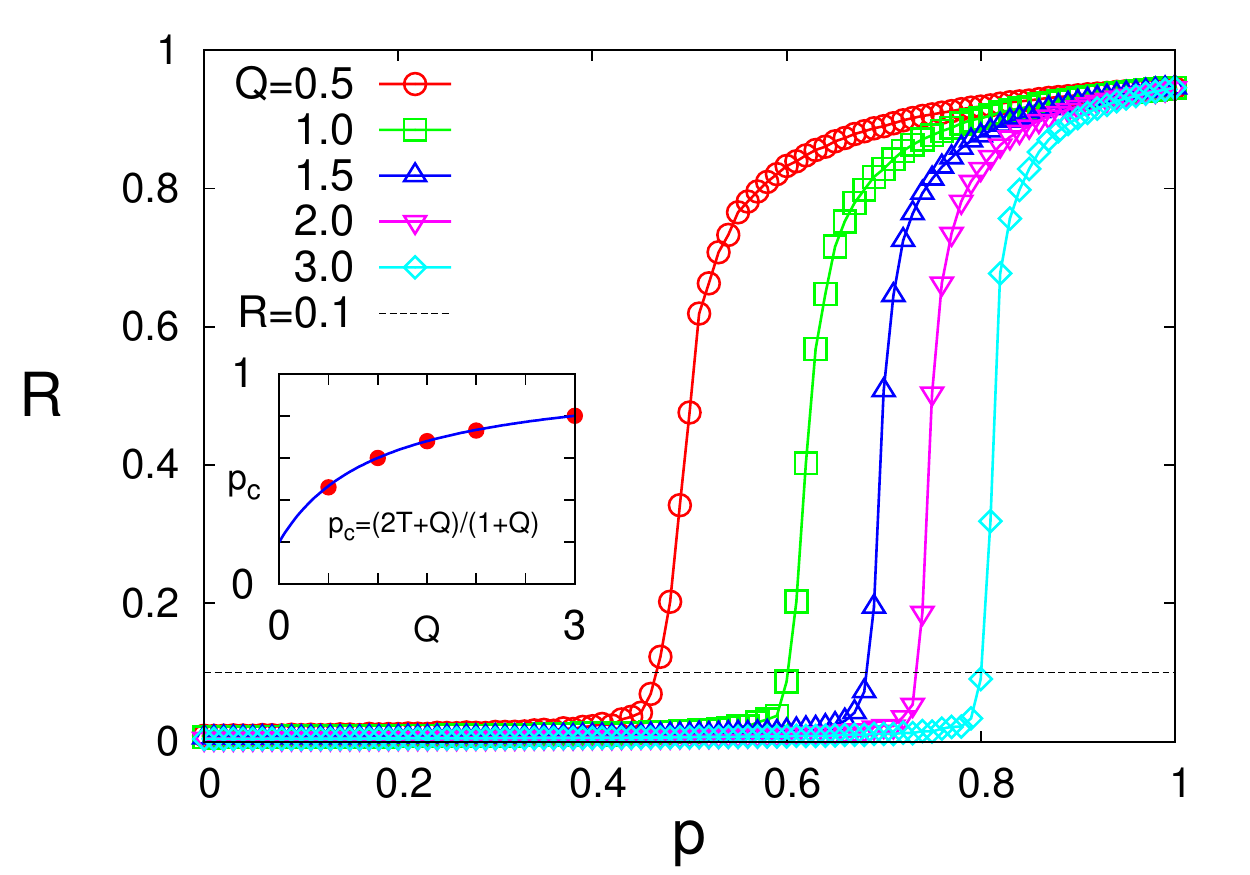}
        \put(0,65){\color{black}(b)}
\end{overpic}
\caption{(Color Online) 
\textcolor{black}{
(a) Steady state behavior of $R$ shown (a) as a function of $p$ with varying values of $T$, for $Q=1$ fixed;
and (b) shown as a function of $p$ with varying values of $Q$, for $T=0.1$ fixed.  
The system size is $N=3200$  and measurements for the order parameter $R$ were averaged over 10 realizations.
Insets: Measurements of $p_c$ vs. $T$ or $Q$, respectively (red dots, see text), which follow the mean-field prediction $p_c=(2T+ Q)/(1+Q)$ (solid blue) from Eq.~\eqref{eq:pc_incoherent_Tgg0}.
}
}
\label{fig:R_p_Q_stochastic}
\end{figure}

To address these two questions, we performed numerical simulations of Eq.~\eqref{eq:model} and measured 
the order parameter $R$ in (\ref{eq:R}) while  varying system parameters.
We first varied the noise level $T>0$ while keeping the coupling ratio $Q=1$ fixed. This revealed features of a phase transition of second-order, as shown in Fig.~\ref{fig:R_p_Q_stochastic}. As temperature increases, so does the critical threshold $p_c$ for the onset of synchrony increase.
Next, we varied $p$ and $Q$ while keeping the temperature fixed at $T=0.1$. The resulting behavior of $R$ is shown in Fig.~\ref{fig:R_p_Q_stochastic}.
For the incoherent state ($R=0$) arising for $p < p_c$, we found that the averaged order parameter $R$ approaches zero as $N^{-1/2}$; for the partially coherent state arising for $p > p_c$, the order parameter $R$ approaches a saturated value $0<R<1$ as $N$ increases, such that any visible dependence on system size is not noticeable once the number of oscillators exceeds $N = 200$. The presence of thermal noise changes the nature of the transition from the incoherent ($R=0$) to the partially coherent state ($R>0$): for $T>0$, the system exhibits a \emph{continuous transition} between the two phases of second-order type, thus contrasting the first-order like transition found for the deterministic case with $T=0$.

To numerically estimate the critical threshold $p_c$  for this phase transition, we detected the value of $p$ for which the time-averaged order parameter exceeds the value of $R = 0.1$, i.e., we used the numerical threshold $\epsilon=0.1$ to identify the incoherent state.
%
(Note that the critical threshold $p_c$ depends on the value $\epsilon$, i.e., larger value of $p_c$ result from larger values of $\epsilon$.) Regardless, the critical threshold values $p_c$ measured for $T>0$ exceed the values of those observed for $T=0$. 
Currently, we lack a bifurcation theory to determine the value of $p_c$ depending on the strength of noise, $T$, is lacking and is left for future study.

The inset of Fig.~\ref{fig:R_p_Q_stochastic} shows the behavior of $p_c$ in dependence of the coupling ratio $Q$ 
while keeping the temperature fixed at $T=0.1$.  
We observed that the threshold $p_c$ increased as temperature, $T$, increased. This is plausible, considering that larger thermal noise renders the synchronization of the system harder to achieve.
Furthermore, increasing the value of coupling ratio $Q$ implies augmenting the dominance of the negative coupling. Thus, the system becomes harder to synchronize, and consequently, the critical threshold, $p_c$, becomes larger.

\subsection{\textcolor{black}{Stability analysis for the incoherent branch in the mean-field limit}}
We apply recent results for a mean-field theory of the Kuramoto model valid for coupling networks with heterogeneous connectivity and coupling strengths, based on the theory of graphons~\cite{Medvedev2014,ChibaMedvedev2016,KaliuzhnyiVerbovetskyiMedvedev} or graphops~\cite{BackhausSzegedy2022,Kuehn2020}.
The basic idea is that we may describe the phase evolution of oscillators in the mean-field limit, $N\rightarrow\infty$, via a density $\rho(t,\cdot)$ that evolves according to a transport equation, the Vlasov-Focker-Planck (VFPE) equation. This idea is orginally formulated for complete graphs with uniform coupling strengths (full graphs), see for instance Refs.~\onlinecite{Strogatz2000,BickMartens2020,Gupta_Campa_Ruffo_2018}. 

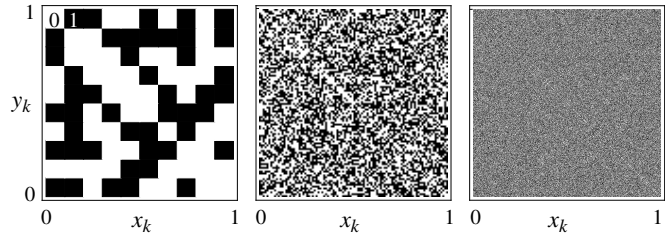
\begin{figure}[htp!]
    \begin{overpic}[width=\columnwidth,percent]{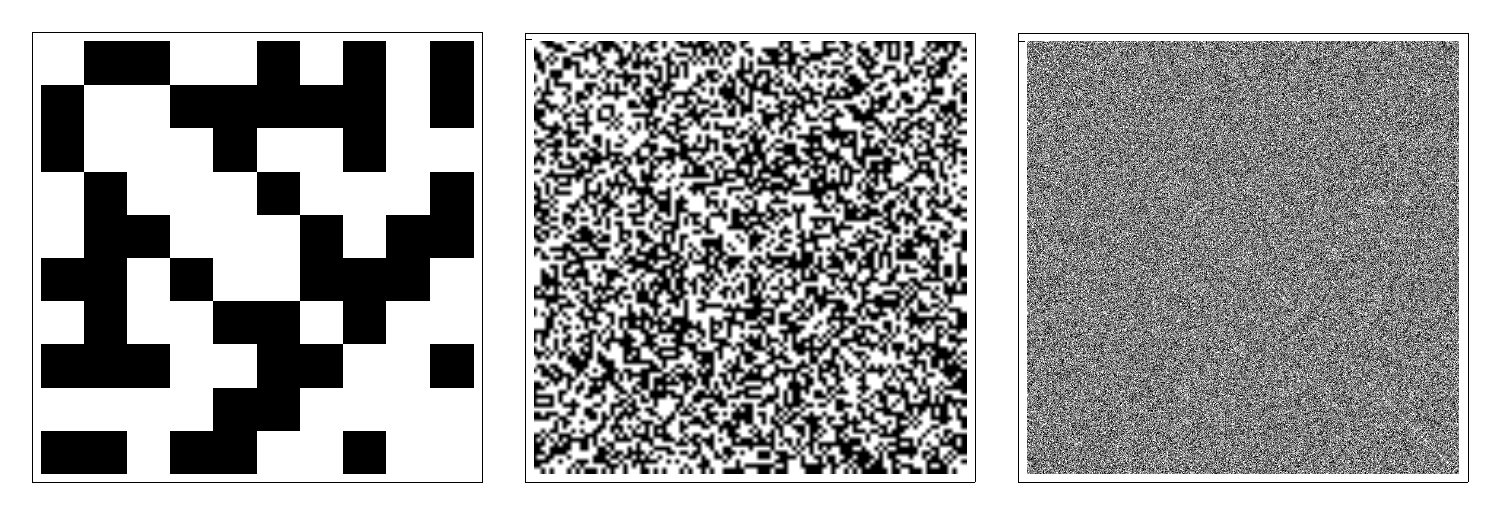}
        \put(-2.5,16){\color{black}$y_k$}
        \put(16,-2.5){\color{black}$x_k$}
        \put(48,-2.5){\color{black}$x_k$}
        \put(80,-2.5){\color{black}$x_k$}
        \put(-0.5,1.5){\color{black}\footnotesize0}
        \put(-0.5,29.5){\color{black}\footnotesize1}
        \put(2,-2){\color{black}\footnotesize0}
        \put(31,-2){\color{black}\footnotesize1}
        \put(35,-2){\color{black}\footnotesize0}
        \put(64,-2){\color{black}\footnotesize1}
        \put(68,-2){\color{black}\footnotesize0}
        \put(96,-2){\color{black}\footnotesize1}
        \put(6.2,28.4){\color{white}\footnotesize1}
        \put(3.2,28.4){\color{black}\footnotesize0}
    \end{overpic}
\caption{Graph sequence $W^{(N),\text{ER}}$for $p=0.5, Q=0.2$ approximating the graphon $W^\text{ER}(x,y)$ with $N=10,100,1000$ (left to right).}
\label{fig:graph_sequence}
\end{figure}

\paragraph{Mean-field theory for heterogeneous networks (graphs).} We briefly outline how such a mean-field theory may be extended to describe a heterogeneous, or even random, network topologies.
To achieve this, it is instructive to consider how one can describe graphs in the mean-field limit, $N\rightarrow\infty$. Consider sequences of weighted graphs $\Gamma_N$ with increasing oscillator number $N$, as illustrated in Fig.~\ref{fig:graph_sequence}. These sequences may be thought of as an approximation of the description of the graph in the mean-field limit. 
Consider a discretization of $I:=[0,1]$, such that $X_n=\{\xiN_1,\ldots,\xiN_N\}$ and 
$\lim_{N\rightarrow\infty}N^{-1} \sum_{i=1}^N f(\xiN_i) =\int_I f(x)\txtd x$ for all continuous functions $f$ on $I$. The points $\{\xiN_i\}_{i\in[N]}$ are sampled either equidistantly, or randomly and uniformly i.i.d.
Let $W$ be a symmetric and an integrable function on $I^2=[0,1]\times[0,1]$. 
The weighted graph $\Gamma_N=G(V,E)$ is defined by the node set $V(\Gamma_N)=[N]$ and edge set $E(\Gamma_N)=\{(i,j)|W(\xi^{(N)}_i,\xi^{(N)}_j) \neq 0; \,i,j\in [N]\}$. Each edge $(i,j)$ is assigned the weight $W^{(N)}_{ij}=W(\xi^{(N)}_i,\xi^{(N)}_j)$; if, on the other hand, the graph is random, we define a W-random graph via $\mathbb{P}((i,j))=W(\xi^{(N)}_i,\xi^{(N)}_j)$ and 0 otherwise, where the decision for each pair $(i,j)$ is made independently.  The limiting behavior of such sequences is determined by a symmetric measurable function on the unit square $I$ called a \emph{graphon}. 
The graphon is a Fredholm integral operator defined by $\mathcal{W}:L^2(I)\rightarrow L^2(I)$,
\begin{align}\label{eq:graphon_defn}
 \mathcal{W}[f](x) = \int_I W(x,y)f(y)\,\txtd y.
\end{align}
%
In the case of an ER graph\eqref{eq:Kmatrix}, shown in Fig.~\ref{fig:graph_sequence}, this sequence converges to a graphon with $W^\text{ER}(x,y)=p$, as shown in Ref.~\onlinecite{ChibaMedvedev2016}. Note that this result represents the exact mean-field limit for $\AER$ in Eq.~\eqref{eq:ER_mf}.

Using graphons it is possible to state a VFPE for a heterogeneous (random) graph structure in the mean-field limit.
For simplicity, consider the deterministic case with a phase density $\rho(t,\theta,x)$. 
The density then evolves according to the transport equation,
\begin{subequations}
\begin{align}\label{eq:transporteqnA}
 \frac{\partial }{\partial t} \rho(t,\theta,x) + 
 \frac{\partial }{\partial \theta} (\rho(t,\theta,x)v(t,\theta,x))&=0, 
\end{align}
with
\begin{align}\label{eq:transporteqnB}
    v(t,\theta,x)&=\omega + K\int_I
    \int_{-\pi}^{\pi}W(x,y)\sin{(\phi-\theta)}
    \rho(t,\phi,y)\,\txtd \phi \,\txtd y,
\end{align}
\end{subequations}
with an initial condition $\rho(0,\theta,x)=g(\theta,x)$ sufficiently smooth.
Here, $x$ can be thought of as a variable describing the heterogeneity of the network so that one effectively obtains a family of transport equations. One may take $x$ as a variable in the unit interval $x \in I=[0,1]$, where points in the interval represent node labels in the infinite network limit ~\cite{ChibaMedvedev2016,KaliuzhnyiVerbovetskyiMedvedev,GkogkasKuehn}. 
It can be shown~\cite{ChibaMedvedev2016} that the density $\rho(t,\cdot)$ is a unique weak solution for $t\in[0,T]$ of the initial value problem \eqref{eq:transporteqnA}-\eqref{eq:transporteqnB}, characterized by an absolutely continuous measure $\mu_t$; and similarly, that the solution of the initial value problem for the finite oscillators \eqref{eq:model} generates a family of empirical measures $\mu_t^N$ for $t\in[0,T]$; furthermore, that these empirical and absolutely continuous measures converge as $N\rightarrow\infty$ for $t\in[0,T]$ which implies convergence of the observed dynamics in the mean-field limit.


\paragraph{Synchronization transition for stochastic $XY$ oscillator model.}%
In Ref.~\onlinecite{Gkogkas2022}, the authors develop a more general mean-field theory valid for the stochastic Kuramoto model; this means in particular that \eqref{eq:transporteqnA} has an additional diffusive term that takes into account fluctuations of strength $T$.
\footnote{More precisely, 
Ref.~\cite{Gkogkas2022} develops a mean-field theory for the stochastic Kuramoto model using graphops (graph 
operators)~\cite{Kuehn2020}, which  generalize graphons (applicable to dense graph structures) and graphings (applicable sparse graph structures); see 
Ref.~\cite{BackhausSzegedy2022} for an in-depth discussion. 
Since our coupling in Eq.~(\ref{eq:Kmatrix}) represents a dense
graph (except for the case where $Q=0$ and $p\rightarrow 0$),
graphons are an adequate description for our analysis.
}
The authors carried out a linear stability analysis for the incoherent solution of the stochastic Kuramoto model with $T>0$. This is performed via linearizing the corresponding VFPE, resulting in an amplitude equation describing the growth (or decay) of perturbations around the incoherent branch. For the special case of the standard Kuramoto model with uniform coupling, $K_{ij}=K$ for all $i,j\in[N]$, and identical frequencies, one obtains the critical coupling for the incoherence-coherence transition,
\begin{align}\label{eq:Kc}
 K_c &= 2T,\
\end{align}
a result that coincides with the classical result by Sakaguchi~\cite{Sakaguchi1988}, and later by Strogatz and Mirollo~\cite{StrogatzMirollo1991}.


The coupling in \eqref{eq:Kmatrix} can be expressed in terms of ER networks --- accordingly, the corresponding graphon valid for the mean-field limit ($N\rightarrow\infty$) is also a linear combination of a constant and the graphon $W^\text{ER}$ for the ER network,
\begin{align}\label{eq:graphon}
    \begin{split}
        W_{p,Q}(x,y) &= 
        -Q + (1+Q) W^\text{ER}(x,y)\\
            &= 
            -Q + (1+Q) p.\
    \end{split}
\end{align}
The last equation follows as the graphon for the ER network has been derived in Ref.~\onlinecite{ChibaMedvedev2016} to  be $W^\text{ER}(x,y) = p$. 
Note that in this case, random sampling of the (finite) graph $\Gamma_N$ must be done by modifying the prescription for the W-random graph introduced further above, i.e., each edge $(i,j)$ is attributed the weight via $\mathbb{P}((i,j))=W_{p,Q}(\xiN_i,\xiN_j)$ and $W_{0,Q}$ otherwise.


Since we chose $K_p=1$ (and $K_n=-Q$) without loss of generality, and the graphon \eqref{eq:graphon} is a constant that can be placed outside the integral in \eqref{eq:transporteqnB}, we effectively have an effective coupling strength $K_\text{eff}=-Q+(1+Q)p$. Letting this effective coupling be critical, $K_\text{eff}=K_c$, we obtain from \eqref{eq:Kc} a
condition for the change of stability of the incoherent branch,
\begin{align}\label{eq:pc_incoherent_Tgg0}
    p_c &= \frac{2T+Q}{1+Q}.
\end{align}
The simulations and measurements of the threshold for the incoherence-coherence transition shown in Fig.~\ref{fig:R_p_Q_stochastic} show an excellent agreement with \eqref{eq:pc_incoherent_Tgg0}.
Note that our previous result in  \eqref{eq:pc_incoherent_T0}  is retrieved as $T\rightarrow 0$.
Finally, the same result for the critical threshold could also be obtained by calculation of the eigenvalues of the graphon~\cite{ChibaMedvedev2016,Gkogkas2022}; however, the above solution is more direct.

\section{Summary and Discussion}
\paragraph{Summary.}
We considered a population of all-to-all coupled oscillators subject to quenched and temporal disorder and explored its collective behavior. The quenched disorder arises due to (symmetric) random coupling interactions with either positive or negative value, and the temporal disorder due to additive thermal noise. For the deterministic case ($T=0$) with \emph{quenched disorder} only, we found that the system displays a discontinuous first-order like phase transition from the incoherent to the fully coherent state.  We provided theoretical arguments for the nature of this transition and explained its critical threshold value $p_c$  for the thermodynamic limit. We confirmed the critical threshold $p_c$ rigorously by providing a stability argument and performing a linear stability analysis of the fully coherent state ($R=1$).
Vice versa, for the noisy case ($T>0$) with additional \emph{temporal disorder}, we found that the system displays a continuous second-order phase transition from the incoherent to the partially coherent state ($0<R<1$). Numerically, we found that the corresponding critical threshold $p_c$ grows with increasing $Q$ and $T$, and we determined an exact formula for the incoherence-coherence transition based on an exact mean field theory~\cite{Gkogkas2022} and a linear stability analysis for the incoherent branch. The formula for this threshold reduces to the results we obtained for $T=0$.


\vspace{5em}

\paragraph{Relation to other studies.} 
A number of studies concerned the collective dynamics in networks of coupled oscillators characterized by heterogeneous properties and interactions. Such models may have contiguous properties such that oscillators form subpopulations or communities with identical properties, giving rise to multimodal frequency distributions or non-uniform coupling interactions demarcating subpopulations of strongly interacting oscillators ~\cite{Abrams2008,Martens2009,MartensBickPanaggio2016,Pietras2018,BickMartens2020,Yamakou2020}. Another type of heterogeneity arises when these properties are non-contiguous, including a variety of oscillator models with heterogeneous positive and negative coupling.
For instance, our model is related to the important class of {\it oscillator glass models}~\cite{Daido1987,Daido1992,Daido2000,Stiller1998,Stiller2000,Ottino2018} which distinguish themselves from \eqref{eq:model} in that they assume distributed natural frequencies and normally distributed coupling strengths $K_{ij}$. A special case represent the {\it $XY$ spin glass models} for identical oscillators with $\omega_i=0$, see Ref.~\onlinecite{SherringtonScott1975}. To render these problems analytically tractable, most studies assumed separable disorder in the coupling, e.g., $K_{ij}=J s_i s_j$ with $s_{i(j)}=\pm 1$ and $J$ a constant. In this case, the spin-glass phase is found to exist in the regime of the incoherent state where the $XY$ spins are fully randomized~\cite{SherringtonScott1975}. For a review on such types of models, see also Ref.~\onlinecite{Acebron2005}.
A simpler coupling scheme arises in systems with oscillator nodes have \emph{uniform input weights}, i.e., $K_{ij}=K_i$ for all $i\in[N]$~\cite{KoErmentrout2008,Laing2009}. Hong and Strogatz~\cite{HongPRL2011,HongPRE2011} studied this case where coupling strengths are either value $K_i=K_p>$ or $K_i=K_n<0$. The oscillators was found to undergo a phase transition from the incoherent ($R=0$) to the partially coherent state ($0<R<1$) when natural frequencies were normally distributed; vice versa, if natural frequencies were identical, oscillators were found to display a phase transition from an incoherent state to a state exhibiting a traveling wave~\cite{HongPRL2011}. 
Alternatively, one may consider \emph{uniform output weights} ($K_{ij}=K_j$ for all $j\in[N]$); for this case it was found that in the presence of heterogeneous frequencies, the system displays conventional mean-field behavior with a second-order phase transition like that found in the original Kuramoto model~\cite{Hong2012}. 
An intriguing extension of coupled oscillator models is the swarmalator model~\cite{OKeeffe2017}, where oscillators are allowed to move on a spatial domain, where their respective level of synchronization influences their motion while their distance modulates their respective coupling strengths. The collective behavior resulting from such a model with the same type of quenched random disorder studied here has recently been investigated~\cite{HongYoeLee2021}. 

\paragraph{Universal threshold.}

Remarkably, the critical threshold $p_c$ (Eq.~(\ref{eq:pc_zeroT})) for quenched disorder ($T=0$) is identical to the one observed for the systems mentioned above, i.e., including uniform input weights~\cite{HongPRL2011} (see also~\cite{HongMartens2021} with $Q=1$ and system parameter $\gamma=0$ therein), uniform output weights~\cite{Hong2012}, as well as the model of identical swarmalators~\cite{HongYoeLee2021} with quenched random disorder.
Indeed, upon further inspection, the universality of the threshold $p_c$ for $T=0$ seems fairly evident from the energetic argument we made. Specifically, the energy for the completely incoherent state is $\mathcal{H}=0$ in the thermodynamic limit $N \rightarrow \infty$, while it is negative for the fully coherent state, $\mathcal{H}<0$. Assuming that the energy density changes continuously, we may identify the critical state as the point where the energy density of the two phases become equal. As we have seen, this critical transition is closely related with the mean value of the coupling disorder, given by
\begin{equation}\label{eq:condition}
    \langle K_{ij} \rangle = K_p \cdot p + K_n \cdot (1-p)
\end{equation}
the value of which also crosses zero at criticality since $\mathcal{E}_\text{coh}=-\langle K_{ij}\rangle$. This argument yields the critical threshold $p_c = Q/(1+Q)$,
and does not depend on the specific structure of the coupling matrix $K_{ij}$ (i.e., the ordering in terms of the values $K_n$ and $K_p$), but rather only on the specific number of entries with negative or positive coupling. Note that the general coupling matrix $K_{ij}$ includes all cases discussed here including uniform input ($K_{ij}=K_i$) or output ($K_{ij}=K_j$). Thus, this explains the universal nature of the critical threshold for the various quenched types of random disorder in the coupling strength discussed here.
However, while this universality of the critical threshold holds when thermal noise is absent ($T=0$), our numerical simulations indicated that this universality breaks down when thermal noise is present ($T>0$). 

\paragraph{Outlook.}
\textcolor{black}{
Furthermore, while most classical studies assumed separable random disorder for the coupling interactions, we considered non-separable coupling interactions. 
For future research, it would be interesting to investigate whether the critical threshold $p_c$ for the noisy case ($T>0$) also has a universal character as the zero-noise case ($T=0$), in terms of various types of quenched disorder in the coupling. 
We leave these questions for a future study.}    

\section{Acknowledgements}
We would like to thank Prof.~Hyunggyu Park and Kangmo Yeo for useful discussions in the early stage of the study; and Prof. Jaegon Um and Prof. Christian Kuehn for helpful discussions regarding the stability analysis, Dr. C. Bick regarding valuable comments on the manuscript. This research was supported by the NRF Grant No.~2021R1A2B5B01001951~(H.H).

\section{Data Availability Statement}
Data sharing is not applicable to this article as no new data were created or analyzed in this study.

\bibliographystyle{unsrt}

\def\tb{\textbackslash}
 
\end{document}